\documentclass[twocolumn,aps,amsmath,amssymb,floatfix,superscriptaddress,prb,longbibliography]{revtex4-1}
\usepackage{graphicx}% Include figure files
\usepackage{dcolumn}% Align table columns on decimal point
\usepackage{bm}% bold math
\UseRawInputEncoding
\usepackage{xspace}
\usepackage{hyperref}
\usepackage{color}
\usepackage[flushleft]{threeparttable}
\def\new{\color{black}}
\def\yzgo{YbZnGaO$_4$\xspace}
\def\ymgo{YbMgGaO$_4$\xspace}
\begin{document}

\title{Disorder-induced broadening of the spin waves in a triangular-lattice quantum-spin-liquid candidate YbZnGaO$_4$}

\author{Zhen~Ma}
\altaffiliation{These authors contributed equally to this work.}
\affiliation{Institute for Advanced Materials, Hubei Normal University, Huangshi 435002, China}
\affiliation{National Laboratory of Solid State Microstructures and Department of Physics, Nanjing University, Nanjing 210093, China}
\author{Zhao-Yang~Dong}
\altaffiliation{These authors contributed equally to this work.}
\affiliation{Department of Applied Physics, Nanjing University of Science and Technology, Nanjing 210094, China}
\author{Jinghui~Wang}
\affiliation{National Laboratory of Solid State Microstructures and Department of Physics, Nanjing University, Nanjing 210093, China}
\affiliation{School of Physical Science and Technology and ShanghaiTech Laboratory for Topological Physics, ShanghaiTech University, Shanghai 200031, China}
\author{Shuhan~Zheng}
\affiliation{Institute for Advanced Materials, Hubei Normal University, Huangshi 435002, China}
\affiliation{National Laboratory of Solid State Microstructures and Department of Physics, Nanjing University, Nanjing 210093, China}
\author{Kejing~Ran}
\affiliation{National Laboratory of Solid State Microstructures and Department of Physics, Nanjing University, Nanjing 210093, China}
\affiliation{School of Physical Science and Technology and ShanghaiTech Laboratory for Topological Physics, ShanghaiTech University, Shanghai 200031, China}
\author{Song~Bao}
\author{Zhengwei~Cai}
\author{Yanyan~Shangguan}
\affiliation{National Laboratory of Solid State Microstructures and Department of Physics, Nanjing University, Nanjing 210093, China}
\author{Wei~Wang}
\affiliation{School of Science, Nanjing University of Posts and Telecommunications, Nanjing 210023, China}
\author{M.~Boehm}
\author{P.~Steffens}
\author{L.-P.~Regnault}
\affiliation{Institut Laue-Langevin, 71 Avenue des Martyrs, 38042 Grenoble Cedex 9, France}
\author{Xiao~Wang}
\author{Yixi~Su}
\affiliation{J\"ulich Centre for Neutron Science (JCNS) at Heinz Maier-Leibnitz Zentrum (MLZ), Forschungszentrum J\"ulich, Lichtenbergstrasse 1, D-85747 Garching, Germany}
\author{Shun-Li~Yu}
\email{slyu@nju.edu.cn}
\author{Jun-Ming~Liu}
\author{Jian-Xin~Li}
\email{jxli@nju.edu.cn}
\author{Jinsheng Wen}
\email{jwen@nju.edu.cn}
\affiliation{National Laboratory of Solid State Microstructures and Department of Physics, Nanjing University, Nanjing 210093, China}
\affiliation{Collaborative Innovation Center of Advanced Microstructures, Nanjing University, Nanjing 210093, China}

%\date{\today}

\begin{abstract}
Disorder is important in the study of quantum spin liquids, but its role on the spin dynamics remains elusive. Here, we explore the disorder effect by investigating the magnetic-field dependence of the low-energy magnetic excitations in a triangular-lattice frustrated magnet YbZnGaO$_4$ with inelastic neutron scattering. With an intermediate field of 2.5~T applied along the $c$-axis, the broad continuum at zero field becomes more smeared both in energy and momentum. With a field up to 10~T, which fully polarizes the magnetic moments, we observe clear spin-wave excitations with a gap of $\sim$1.4~meV comparable to the bandwidth. However, the spectra are significantly broadened. The excitation spectra both at zero and high fields can be reproduced by performing classical Monte Carlo simulations which take into account the disorder effect arising from the random site mixing of nonmagnetic Zn$^{2+}$ and Ga$^{3+}$ ions. These results elucidate the critical role of disorder in broadening the magnetic excitation spectra and mimicking the spin-liquid features in frustrated quantum magnets.
\end{abstract}

\maketitle

\section{Introduction}

Quantum spin liquids (QSLs) represent a nontrivial state of matter in which spins are highly correlated but fluctuate quantum mechanically and thus do not form a long-range magnetic order even in the zero-temperature limit\cite{Anderson1973153,nature464_199,
RevModPhys.89.025003,Broholmeaay0668}. Achieving such a state has been a long-sought goal as they are believed to host fractional excitations and emergent gauge structures that can find applications in quantum computation\cite{Kitaev20032,Barkeshli722,0034-4885-80-1-016502}.
Magnetic frustration is an important ingredient for a QSL. Besides the exchange frustration on a single site resulting from anisotropic interactions such as the bond-dependent Kitaev interactions defined on the honeycomb lattice\cite{aop321_2,prl102_017205,0953-8984-29-49-493002,
PhysRevLett.118.107203,npjqm4_12,nrp1_264}, another prototypical type is the geometrical frustration which describes the situation that magnetic exchange interactions cannot be satisfied simultaneously among different lattice sites due to the geometrical constraint\cite{Anderson1973153,arms24_453,nature464_199}.
By introducing geometrical frustration into a low-spin (such as $S=1/2$) system to enhance quantum fluctuations, a static magnetic order is avoided \cite{Anderson1973153,arms24_453,nature464_199}. Consequently, materials with effective spin-1/2 on two-dimensional triangular or kagome lattice with strong geometrical frustration are regarded as compelling platforms to host a QSL state\cite{PhysRevLett.91.107001,PhysRevLett.95.177001,
np4_459,np5_44,nc9_307,np6_673,Yamashita1246,nc2_275,
Liu_2018,PhysRevB.98.220409,np15_1058,PhysRevB.100.144432,
PhysRevB.100.241116,prl98_107204,nature492_406,
RevModPhys.88.041002,ZiliFeng:77502,PhysRevB.102.224415}. A QSL phase has also been proposed on a square lattice where strongly competing interactions along different exchange paths are present\cite{PhysRevB.98.241109,nc9_1085,PhysRevLett.121.107202,PhysRevB.98.134410,PhysRevB.102.014417}.

Along this line, a triangular-lattice compound \ymgo has attracted tremendous attention due to the possible realization of a gapless QSL state\cite{sr5_16419,prl115_167203,nature540_559,np13_117,
PhysRevLett.117.097201,PhysRevLett.118.107202,nc8_15814,
PhysRevLett.122.137201}. Various experimental techniques such as heat capacity\cite{sr5_16419}, electron spin resonance\cite{PhysRevLett.117.097201}, and muon spin relaxation ($\mu$SR)\cite{PhysRevLett.117.097201,PhysRevB.102.014428}, show neither signature of long-range magnetic order nor symmetry breaking down to several tens of millikelvin. In particular, a broad continuum of magnetic excitation spectra observed by inelastic neutron scattering (INS) provides strong evidence for a QSL state with fractionalized spinon excitations\cite{nature540_559,np13_117,nc9_4138}. However, thermal conductivity $(\kappa)$ study shows that the gapless magnetic excitations do not contribute to $\kappa$, challenging the idea of $U(1)$ gapless QSL with itinerant spinons in this material\cite{PhysRevLett.117.267202}. In a previous study\cite{PhysRevLett.120.087201}, we replaced the nonmagnetic Mg with Zn, and by overcoming the volatile problem caused by ZnO, we successfully grew large-size high-quality single crystals for an isostructural triangular-lattice compound \yzgo. With comprehensive measurements including dc susceptibility, ultralow-temperature specific heat, and INS, we found that the features were consistent with a gapless QSL. On the other hand, our further ultralow-temperature thermal conductivity measurements revealed that there was no linear term contributed by fermions such as spinons. Subsequent ultralow-temperature ac susceptibility measurements captured a broad peak with frequency dependence both in \ymgo and \yzgo\cite{PhysRevLett.120.087201,nc12_4949}. {\new Furthermore, nonmagnetic Mg$^{2+}$/Zn$^{2+}$ and Ga$^{3+}$ ions are randomly distributed on the equivalent sites on the triangular lattice with a 1:1 occupancy\cite{np13_117,PhysRevLett.120.087201}. Considering the possible disordered exchange couplings caused by the random charge environment as a result of the site mixing
and the small exchange coupling\cite{np13_117,nc8_15814}}, these results were interpreted as originating from a disorder-induced spin-glass ground state\cite{PhysRevLett.120.087201}. Theoretical proposals to describe the ground state for these materials are also divergent, including $U(1)$ gapless QSL with a large spinon Fermi surface\cite{PhysRevB.96.054445,PhysRevB.96.075105}, stripe-ordered phase\cite{PhysRevLett.119.157201,PhysRevB.95.165110,
PhysRevB.97.184413,PhysRevX.9.021017,PhysRevB.103.104421}, and random-singlet state\cite{PhysRevX.8.031028,PhysRevB.103.205122,
nc9_4367}. By now, there has been no consensus on the exact ground state of \ymgo and \yzgo. The central issue of this debate is whether the system is susceptible to disorder due to the random distribution of the nonmagnetic ions\cite{npjqm4_12,Broholmeaay0668}. Although many efforts have been made to elucidate this point\cite{np13_117,PhysRevLett.117.267202,arXiv:1705.05699,
PhysRevLett.119.157201,PhysRevB.97.125105,PhysRevB.97.184413,
PhysRevLett.120.207203,PhysRevX.8.031001,PhysRevX.8.031028,
PhysRevX.9.021017,nc9_4138,PhysRevB.103.104421,PhysRevB.102.104433,
PhysRevB.103.205122,PhysRevResearch.2.023191,PhysRevB.102.014428,
PhysRevResearch.3.033050,npjQM6_78,nc12_4949}, the role of disorder on the spin dynamics is still controversial. Especially, because of the limit on the sample availability for \yzgo, research on this compound is rather scarce\cite{PhysRevLett.120.087201,npjQM6_78}.

\begin{figure}[htb]
\centerline{\includegraphics[width=3.4in]{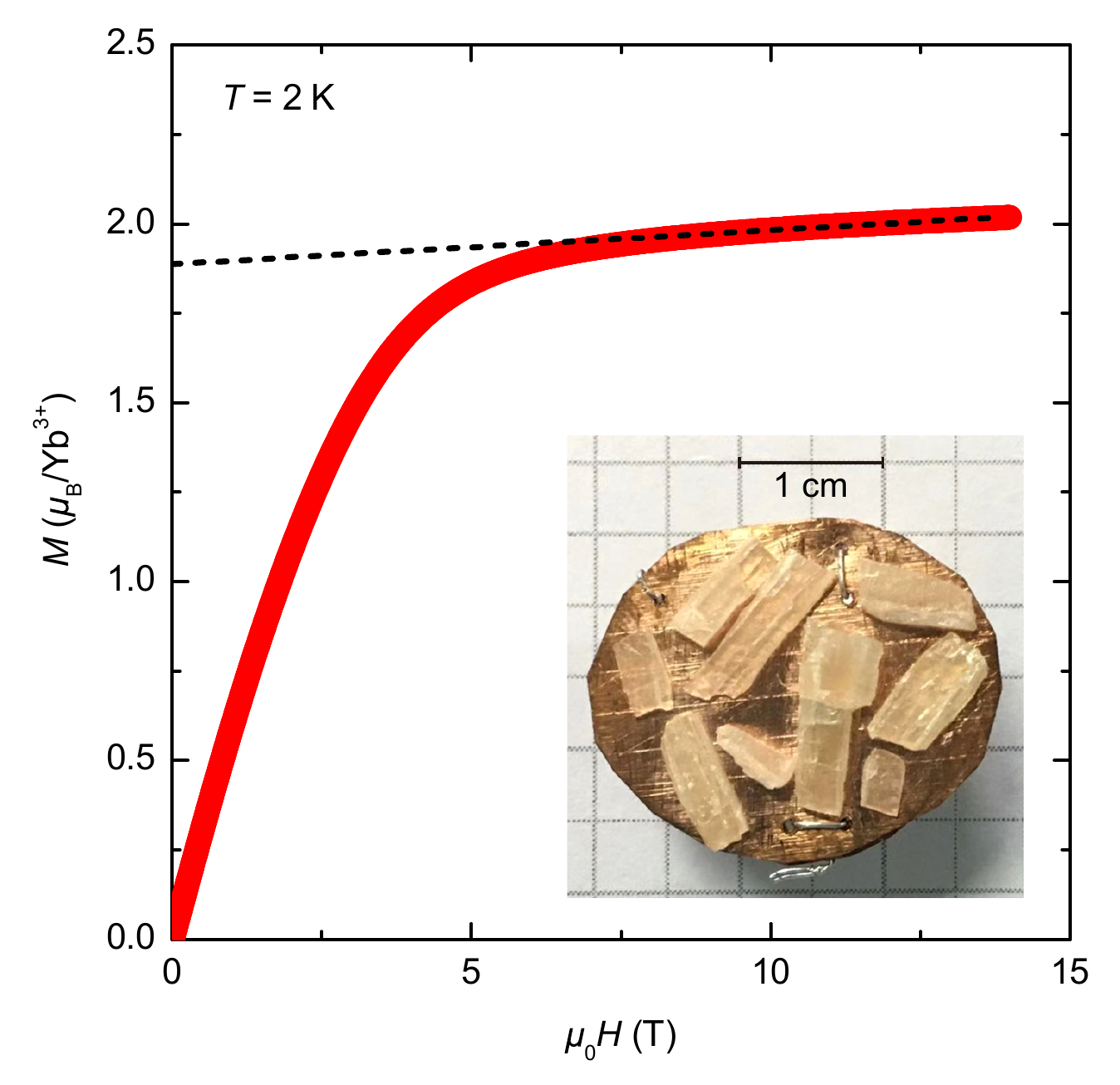}}
\caption{{\bf Magnetization data for a \yzgo single crystal.}
Magnetic-field dependence of the magnetization for a \yzgo single crystal at $T = 2$~K with a field applied parallel to the $c$-axis up to 14~T. The dashed line represents a linear fit to the data for $\mu_{\rm{0}}H~\ge8$~T. The inset shows a photograph of co-aligned 9 pieces of single crystals glued on a copper plate.
\label{fig1}}
\end{figure}

In this work, we carry out INS measurements on single crystals of \yzgo under a $c$-axis magnetic field to examine the field evolution of the magnetic excitations. The magnetization process with magnetic field parallel to the crystalline $c$-axis is shown in Fig.~\ref{fig1}. The magnetization curve increases progressively with field below around 3~T, followed by a smooth transition to a nearly saturated regime above 8~T. As a consequence, a moderate field of 2.5~T redistributes the spectral weight of the broad continuum at zero field in a more uniform fashion in the energy-momentum space. With a field of 10~T, which drives the system into a fully polarized state, we observe clear spin-wave excitations. The excitation spectra are, however, broadened substantially in contrast to well-defined spin waves arising from the clean ferromagnetic state. These features can be simulated by classical Monte Carlo calculations by including the disorder effect resulting from the random mixing of Zn$^{2+}$ and Ga$^{3+}$ ions. These results indicate the presence of significant disorder in \yzgo, which plays an important role in the mimicry of the spin-liquid characteristics in frustrated quantum magnetic systems.

\section{Experimental Details}

High-quality single crystals of \yzgo were grown by optical floating-zone technique under high pressure\cite{PhysRevLett.120.087201}. Crystals with a typical size of 10$\times$3$\times$1.5~mm$^3$ for a piece are shown in the inset of Fig.~\ref{fig1}. The magnetization was measured on a 29.4-mg single crystal with a magnetic field applied along the $c$-axis in a Quantum Design physical property measurement system~(PPMS). INS experiments on the single-crystal sample were carried out on ThALES, a cold triple-axis spectrometer at Institut Laue-Langevin (ILL) at Grenoble, France. In the measurements, we used 9 pieces of single crystals weighed 2.3~g in total, coaligned with a neutron Laue diffractometer NLaue located at Heinz Maier-Leibnitz Zentrum~(MLZ) at Garching, Germany. The crystals were mounted onto a copper sample holder with the $c$-axis perpendicular to the horizontal plane and glued tightly with CYTOP, so that the $(H,\,K,\,0)$ plane is the scattering plane (see the inset of Fig.~\ref{fig1}). The half width at half maximum (HWHM) of the rocking scan through the (2, 0, 0) peak with the incident neutron energy of 16.28~meV was 0.41$^{\circ}$, indicating a good coalignment of the single crystals. A dilution refrigerator was equipped, so that it could cool down the sample to 50~mK. We used a fixed-final-energy $(E_{\rm{f}})$ mode with $E_{\rm{f}} = 3.5$~meV. A Be filter was placed after the sample to reduce high-order neutron contaminations. A double-focusing mode without additional collimators was used for both the monochromator and analyzer. Under such conditions, the energy resolution was about 0.08~meV (HWHM). The wave vector $\bm{Q}$ was expressed as ($H,\,K,\,L$) reciprocal lattice unit (r.l.u.) of $(a^{*},\,b^{*},\,c^{*})=(4\pi/\sqrt{3}a,\,4\pi/\sqrt{3}b,\,2\pi/c)$, with $a=3.414(2)$~\AA, and $c=25.140(2)$~\AA.

\section{Results}
\subsection{Magnetic-field dependence of the magnetic excitation spectra}

\begin{figure*}[htb]
\centerline{\includegraphics[width=6.8in]{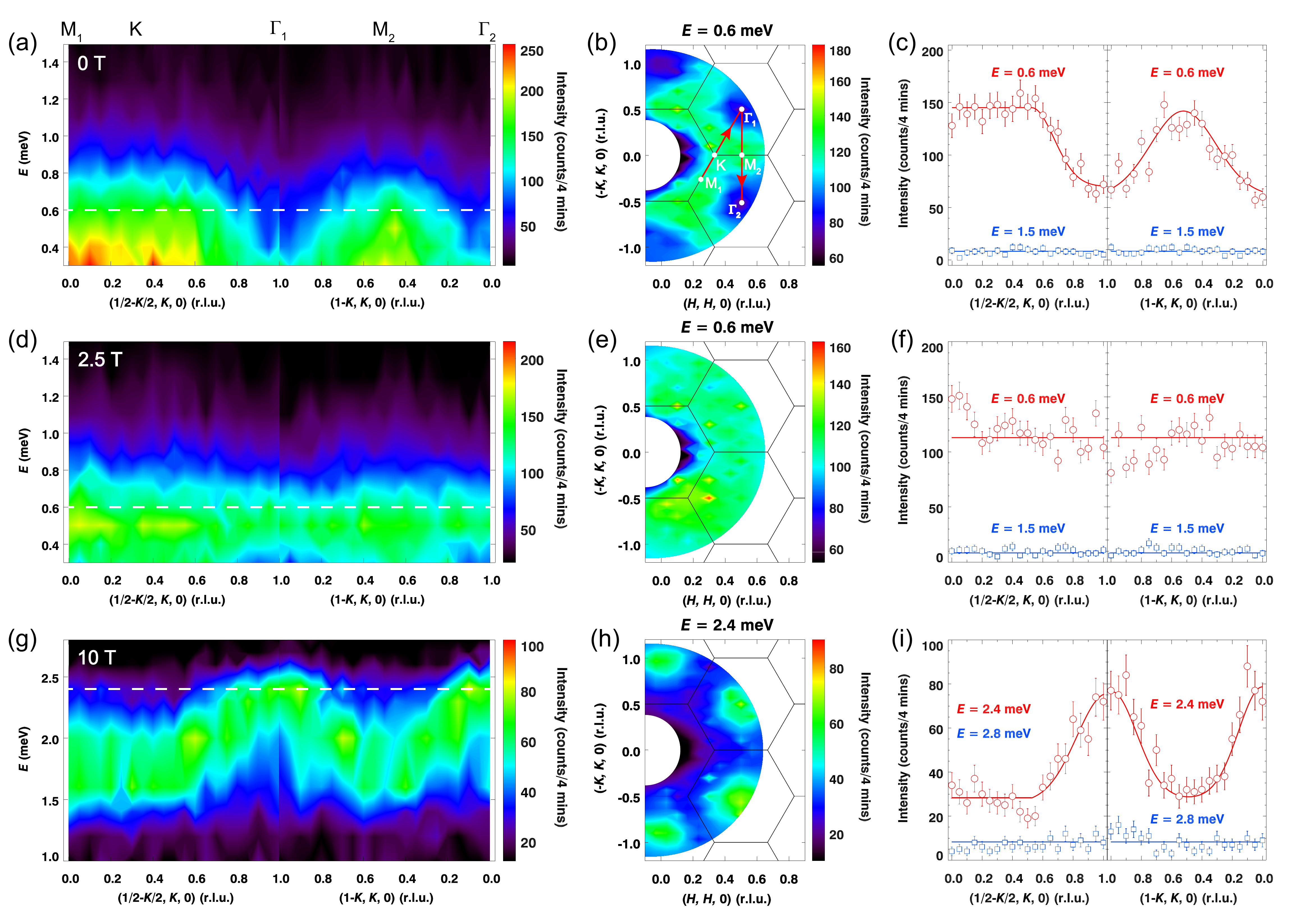}}
\caption{ {\bf Magnetic excitation spectra obtained by INS experiment under different magnetic field strengths at 50~mK.} Magnetic dispersions along high-symmetry directions of M$_1$-K-$\Gamma_1$ and $\Gamma_1$-M$_2$-$\Gamma_2$ as illustrated by the arrows in (b) at 0-T (a), 2.5-T (d), and 10-T (g) fields. The dispersions are obtained by plotting together a series of constant-energy scans displayed in (c), (f), and (i), with an energy interval of 0.1~meV. Dashed lines in (a), (d), and (g) indicate constant-energy scans of 0.6, 0.6, and 2.4~meV, respectively. Solid lines in (c), (f), and (i) are guides to the eye. Contour plots of the INS spectra at $E= 0.6$~meV at 0~T (b), 2.5~T (e), and at $E= 2.4$~meV at 10 T (h). The contour maps are obtained by plotting together a series of constant-energy scans along the $[H,\,0,\,0]$ direction with a step of 0.1~r.l.u. and an interval of 0.1~r.l.u. along the $[0,\,K,\,0]$ direction. Black solid lines represent the Brillouin zone boundaries. Errors represent one standard deviation throughout the paper.
\label{fig2}}
\end{figure*}

We have examined the magnetic-field evolution of the spin excitations and the spectra at 0, 2.5, and 10-T fields with field applied along the $c$-axis are shown in Fig.~\ref{fig2}. Figure~\ref{fig2}(a) displays the dispersions at 0~T along two high-symmetry directions of M$_1$-K-$\Gamma_1$, and $\Gamma_1$-M$_2$-$\Gamma_2$ as illustrated in Fig.~\ref{fig2}(b). We observe continuous magnetic excitations in contrast to well-defined spin-wave excitations\cite{PhysRevLett.113.047202,PhysRevLett.118.107203}. In Fig.~\ref{fig2}(b), we present a contour map with $E= 0.6$~meV at 0~T. Broad diffusive scattering signals along edges of the Brillouin zones are captured. This feature is also completely different from conventional spin-wave excitations whose intensity is mainly concentrated on some specific positions in the reciprocal space\cite{PhysRevLett.113.047202,PhysRevLett.118.107203}. Figure~\ref{fig2}(c) shows two constant-energy scans at $E= 0.6$ and 1.5~meV along the same two high-symmetry paths as in Fig.~\ref{fig2}(a). At $E=0.6$~meV, the scattering intensity is distributed along the edges and the intensity at the high-symmetry M and K points  is almost identical, but drops steeply at the $\Gamma$ points. As for the case of $E=1.5$~meV, magnetic excitations vanish as can be seen in Fig.~\ref{fig2}(c), which represents the upper boundary of magnetic spectra. The broad continuum over the whole energy range measured by INS is consistent with our previous report\cite{PhysRevLett.120.087201}. However, its origin in \yzgo, and similarly in \ymgo, is still hotly debated\cite{nature540_559,np13_117,
PhysRevLett.119.157201,PhysRevB.95.165110,PhysRevB.97.184413,
PhysRevX.9.021017,PhysRevLett.120.087201,
PhysRevX.8.031028,PhysRevResearch.3.033050,npjQM6_78}.

To further understand the magnetic spectra at 0~T and reveal the ground state of \yzgo, we performed INS investigations at finite fields. Figure~\ref{fig2}(d) shows the magnetic dispersions at an intermediate field of 2.5~T that places the sample in the partially-polarized regime. At the first glance the overall continuous excitations captured at 0~T still remains, but some spectral weight around the high-symmetry M and K points shifts to $\Gamma$ points between 0.4 and 0.7~meV. In Fig.~\ref{fig2}(e), we present a contour map at $E=0.6$~meV, and the spectral weight spreads almost uniformly over the Brillouin zone. This phenomenon is more clearly exhibited with two $\bm{Q}$ scans in Fig.~\ref{fig2}(f). In our measured paths, the spin excitation intensity is nearly constant everywhere. Compared with the broad continuous spectra at 0~T, a moderate field of 2.5~T makes the spectral weight distribute more uniformly in the energy-momentum space. The results at the intermediate field are similar to earlier results in \yzgo~(Ref.~\onlinecite{npjQM6_78}).

From the magnetization data displayed in Fig.~\ref{fig1}, we see the compound enters the fully polarized state at 8~T. Therefore, we are able to obtain the magnetic excitation spectra for the ferromagnetic state when performing INS measurements at a field up to 10~T. Magnetic dispersions along the same high-symmetry directions as those at zero and 2.5-T fields are shown in Fig.~\ref{fig2}(g). Clear dispersive spin waves are observed, in contrast to the continuua shown in Fig.~\ref{fig2}(a) and (d). Along M$_1$-K-$\Gamma_1$, the dispersion is nearly flat from M$_1$ to K, and then turns upwards and reaches its maximum at about 2.5~meV. For the direction $\Gamma_1$-M$_2$-$\Gamma_2$, the spectra disperse upwards from M$_2$ and reach the band top at 2.5~meV. A spin gap of $\sim$1.4~meV is also clearly observed. Its value is comparable to that of 1.2~meV in \ymgo at a 9.5-T field\cite{nc9_4138}. We also plot a contour map in the fully polarized state at $E= 2.4$~meV in Fig.~\ref{fig2}(h). It is nearly approaching the band top of the spin-wave excitations. Compared with the case at zero and 2.5-T fields, the scattering patterns are quite different. Now the spectral weight entirely concentrates around the center of the Brillouin zones, so that the intensity at the $\Gamma$ points is much larger than that at the M and K points. It is depicted more clearly in Fig.~\ref{fig2}(i). From Fig.~\ref{fig2}(g)-(i), we can also observe significant broadening of the spin-wave excitation spectra, which is almost comparable to the total bandwidth and well beyond the instrument resolution of 0.08~meV. Similar observations have also been documented in Refs.~\onlinecite{np13_117,nc9_4138,npjQM6_78}. Such broadening
is unexpected for a clean ferromagnetic state. We believe disorder effect should be mainly responsible for this unusual feature as we discuss below.

\subsection{Monte Carlo simulations}

\begin{figure}[htb]
  \centering
  \includegraphics[width=0.98\linewidth]{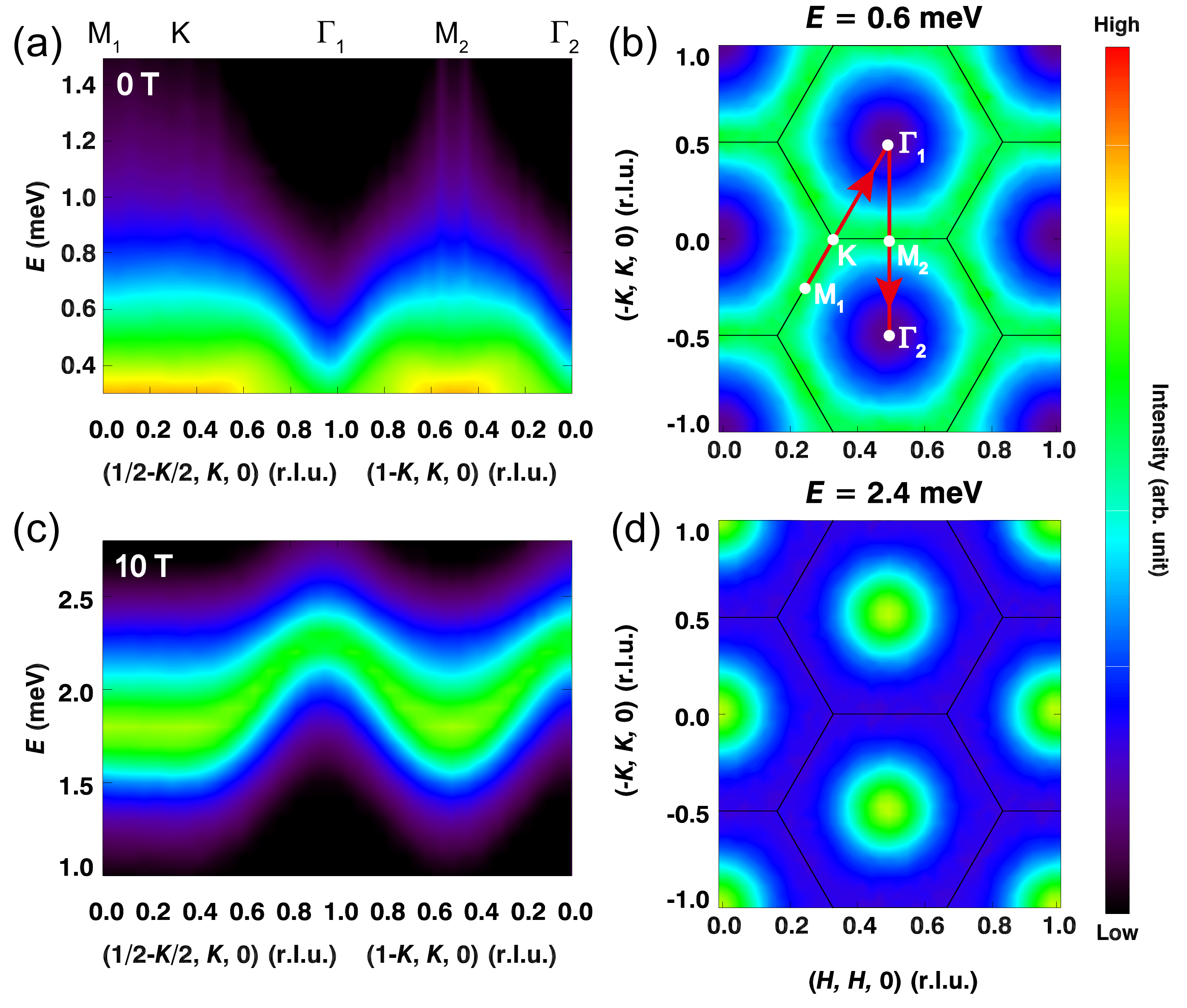}
  \caption{{\bf Results of Monte Carlo simulations.}
  Calculated magnetic dispersions along two high-symmetry routes of M$_1$-K-$\Gamma_1$ and  $\Gamma_1$-M$_2$-$\Gamma_2$ as illustrated by the arrows in (b) at 0-T (a) and 10-T (c) fields. Contour maps of the calculated spectra at $E$ = 0.6~meV at 0-T field (b), and $E= 2.4$~meV at 10-T field (d).} \label{fig3}
\end{figure}

In Ref.~\onlinecite{PhysRevB.102.224415}, we have demonstrated that disorder can induce spin-liquid behaviors in frustrated kagome-lattice compounds Tm$_3$Sb$_3$Zn$_2$O$_{14}$ and Tm$_3$Sb$_3$Mg$_2$O$_{14}$. Particularly, in our previous work on \yzgo~(Ref.~\onlinecite{PhysRevLett.120.087201}), using the linear spin-wave theory, we have pointed out that the spin-liquid-like broad continuous spectra in the material can be well described by an anisotropic spin model with the nearest-neighbor (NN) and next-nearest-neighbor (NNN) exchange interactions after introducing the disorder effect arising from the random mixing of Zn$^{2+}$ and Ga$^{3+}$ into a stripe-ordered phase. Now we also begin with this model and try to reproduce the significantly broadened magnetic excitation spectra at 10-T field which already fully polarizes the moments. {\new We realize that the exact ground state for this material is under debate. Therefore, in order not to be biased by selecting the ground state, we use the Monte Carlo method which can calculate the magnetic state self-consistently with disorder and magnetic field to simulate the magnetic excitation spectra\cite{1999mcms.book.....N,1986TCP.....7.....B,sala2014vacancy,bowman2019role}, rather than using the linear spin-wave theory that needs to define an exact ground state as the starting point. We consider the anisotropic effective spin-wave model proposed in our previous paper\cite{PhysRevLett.120.087201}, and obtain the spin Hamiltonian:}
\begin{eqnarray}
  H=&&\sum_{\langle ij\rangle}[J_{zz}S_i^zS_j^z+J_\pm(S_i^+S_j^-+S_i^-S_j^+)\nonumber\\
  &&+J_{\pm\pm}(\gamma_{ij}S_i^+S_j^++\gamma_{ij}^\ast S_i^-S_j^-)\nonumber\\
  &&+{iJ_{z\pm} \over 2}(\gamma_{ij}^\ast S_i^+S_j^z-\gamma_{ij}S_i^-S_j^z+\langle i\longleftrightarrow j\rangle)]\nonumber\\
  &&+\sum_{\langle\langle ij\rangle\rangle}[J_{2z}S_i^zS_j^z+J_{2\pm}(S_i^+S_j^-+S_i^-S_j^+)]\nonumber\\
  &&-\sum_i[g_\perp(H^xS_i^x+H^yS_i^y)+g_\parallel H^zS_i^z], \label{meanfield}
\end{eqnarray}
where the phase factor $\gamma_{ij}$ is 1, e$^{i2\pi/3}$, and e$^{-i2\pi/3}$ for each of the three directions of the triangular lattice. $\langle ij\rangle$ and $\langle\langle ij\rangle\rangle$ indicate the NN and NNN bonds, respectively. Since the external field we applied is parallel to the $c$-axis, both values of $H^x$ and $H^y$ are equal to zero. {\new To simplify the fittings, some initial constraints are made on the parameters with $J_{z\pm}=2J_{\pm\pm}$ and $J_2=0.1J_1$, based on some reported cases in the literatures\cite{np13_117,PhysRevLett.119.157201,PhysRevResearch.3.033050,PhysRevLett.120.087201}. Herein, $J_2$ is anisotropic as that in our previous work, where it was introduced to stabilize the stripe-order ground state\cite{PhysRevLett.120.087201}. Then the simulating parameters of the Hamiltonian are set to be, $J_\pm=0.66J_{zz}$, $J_{z\pm}=2J_{\pm\pm}=0.13J_{zz}$, $J_{2z}=0.1J_{zz}$, $J_{2\pm}=0.066J_{zz}$, and $J_{zz}=0.14$~meV, which are mainly determined by fitting the energies at high-symmetry M, K and $\Gamma$ points of the spectra at 10~T.} For the Land\'{e} $g$ factors, we use $g_\perp=3.17(4)$ and $g_\parallel=3.82(2)$~(Ref.~\onlinecite{PhysRevLett.120.087201}). As we discuss above, the disorder resulting from the random mixing of the Zn$^{2+}$ and Ga$^{3+}$ is significant in \yzgo. Thus we introduce Gaussian-distributed disorder into the exchange interactions and $g$ factors, with $J_{ij}=J(1+\Delta_{ij})$ and $g_{\parallel i}=g_\parallel(1+\Delta_i)$, where  $\Delta_{ij}$ and $\Delta_i$ satisfy Gaussian distributions. {\new The Gaussian variances are adjusted to simulate the effect of disorder to result in an appropriate broadening of the spectra. During the fitting, a large value of $\Delta_{ij}$ = 0.5 is needed to match the experimental results. A too large $\Delta_i$ will cause an unstable ground state at 0~T, and it is finally determined to be 0.28 that can well reproduce the broadening features. Totally, there are five free parameters used in fitting the experimental spectra.}

To account for the effects of classical fluctuation of disorder, we calculate the ground states by classical Monte Carlo simulations with a standard Metropolis sampling algorithm\cite{1953JChPh..21.1087M}. Then the {\new transverse} dynamical structure factor
\begin{equation}
{\new S(\bm{Q},\omega)}=\frac{1}{N}\sum_{ij}{\rm e}^{\mathrm{i}\bm{Q}(\bm{r_i}-\bm{r_j})}\int_{-\infty}^{\infty}\left\langle S_i^- S_j^+(t)\right\rangle {\rm e}^{-\mathrm{i}\omega t}dt,
\end{equation}
is calculated by Landau-Lifshitz-Gilbert spin dynamics\cite{Gilbert1956,1353448}. The disorder is introduced into the dynamics as Langevin equations,
\begin{equation}
\bm{\dot{S}}_i=\bm{S}_i\times(\mathfrak{F}_i+\mathfrak{F}^\Delta_i), ~\mathfrak{F}_i=-{\partial H_0\over \partial \bm{S}_i}, ~\mathfrak{F}_i^\Delta=-{\partial H_\Delta\over \partial \bm{S}_i},
\end{equation}
where $H_0$ is the Hamiltonian without disorder while $H_\Delta$ is the disorder term of the Hamiltonian.

The dynamical structure factors are calculated by averaging over $1024$ configuration samples on a $32\times32$-point lattice {\new with periodic boundary condition at $50$~mK, and representative results are convergent, as shown in Fig.~\ref{fig3}. We have also tried a smaller system such as a $16\times16$ lattice with 64 samples. Although the spectra are similar with the results shown in Fig.~\ref{fig3}, the calculations are not convergent. For the $16\times16$ lattice, if we increase the sample size to 256, the calculations are convergent and the results have no visible difference from those in in Fig.~\ref{fig3}. These results show that the dependence of our calculation results on the lattice size is not obvious. In the whole simulation process, the broadening caused by the disorder is much larger than the instrument resolution of 0.08~meV, and thus the resolution is not considered.} To assure the validity of the Monte Carlo simulations, we first calculate the zero-field magnetic excitation spectra and the results are shown in Fig.~\ref{fig3}(a) and (b). It is clear that the broad continuous excitation spectra observed experimentally at zero field as shown in Fig.~\ref{fig2}(a) and (b) can be well reproduced by the Monte Carlo simulations. Moreover, the simulations are consistent with our previous calculated results using the linear spin-wave theory by introducing disorder into the stripe order phase\cite{PhysRevLett.120.087201}, which further strengthens the conclusion that in \yzgo disorder plays a significant role in melting the magnetic order. In fact, in our Monte Carlo calculations, before adding disorder, the ground state is also calculated to be a stripe order state, consistent with previous results\cite{PhysRevB.95.165110,PhysRevLett.119.157201,
PhysRevLett.120.087201}.

With this as the starting point, we perform Monte Carlo calculations at a 10-T field, and the resulted magnetic excitation spectra are shown in Fig.~\ref{fig3}(c) and (d). After introducing disorder and a 10-T field into the Hamiltonian, the system evolves self-consistently into a ferromagnetic state. As a consequence, the calculations show clear spin-wave excitations as expected from a magnetic-filed-driven ferromagnetic state. The main features of the spectra, including the gap, band bottom, band top, and the shape of the dispersion are all consistent with the experimental data shown in Fig.~\ref{fig2}(g). The contour map near the band top shows that the spectra weight is concentrated around the $\Gamma$ points, which is also fully consistent with the experimental data shown in Fig.~\ref{fig2}(h). Most importantly, the noticeable broadenings both in the energy and momentum axes in the experimental data, which are unexpected for a ferromagnetic state, can also be nicely reproduced by the calculations. Obviously, the disorder in the exchange interactions and $g$ factors as discussed above are responsible for this feature, thus emphasizing the important role of disorder even in the fully polarized state in this material. 

{\new It is noted that a very recent work reported a phase diagram of \yzgo under fields via linear-spin-wave simulations\cite{npjQM6_78}. They started with a disorder-free XXZ model and obtained the parameters somewhat different from ours. In a zero-field phase diagram of the anisotropic $J_1-J_2$ model, our parameters are deep in the stripe-${yz}$ phase, while the parameters in Ref.~\onlinecite{npjQM6_78} are also in the stripe-$yz$ phase but close to the 120$^\circ$-order state. However, in Ref.~\onlinecite{npjQM6_78}, the calculated spectra with a disorder-free XXZ model were well-defined spin waves dispersing from the M point, distinct from our broad continuous spectra along the boundary of the Brillouin zone as shown in Fig.~\ref{fig3}(a) and (b), obtained by Monte Carlo simulations with disorder. Our simulations are in fact consistent with the experimental observations as shown in Fig.~\ref{fig2}(a) in our work and in Ref.~\onlinecite{npjQM6_78}. This strongly indicates the necessity for the inclusion of disorder, supporting our conclusion. In the high-field state, the calculations in Ref.~\onlinecite{npjQM6_78} also produced sharp spin waves, without broadening features introduced by disorder in our calculations. Even excluding the broadening effect, the dispersion, especially the positions of the band top and bottom in the fully polarized high-field state in Ref.~\onlinecite{npjQM6_78} is also different from ours. The difference is beyond the scope of the field strength and should be resulting from the two different sets of parameters.}

{\new Here, we think it is necessary to clarify some issues about the Monte Carlo simulations: i) With the 10-T experimental spectra, the constraints on the fittings are stronger, yielding more suitable parameters than those of our previous work via linear-spin-wave calculations on the broad continuous data at zero field\cite{PhysRevLett.120.087201}. ii) Although our theoretical calculations can well reproduce the experimental results of 0~T and 10~T, our classical Monte Carlo simulations ignore the quantum fluctuations of disorder, resulting in a premature polarized state at a weak field, such as 2.5~T. This problem was also found in a very recent work via linear-spin-wave simulations, which also ignored the quantum fluctuations of disorder\cite{npjQM6_78}. iii) We also point out that the magnetic excitations in the experiment at 0~T are roughly twice as intense as those at 10~T, while this information cannot be reproduced quantitatively in the calculations.}

\section{Discussions}
Similar to \ymgo, \yzgo as a highly frustrated triangular-lattice compound exhibits many features mimicking those of a QSL\cite{PhysRevLett.120.087201}. Especially, the broad continuous excitation spectra such as those shown in Fig.~\ref{fig2}(a) and (b) have been taken as a smoking gun for QSLs\cite{nature492_406,np12_942,nature540_559,np13_117}. However, as we demonstrate above by Monte Carlo calculations and in our previous work by the linear spin-wave theory\cite{PhysRevLett.120.087201}, such ``continua" can be well reproduced by introducing disorder into the stripe order ground state. More importantly, if there were no disorder, the spin waves in the fully polarized state will be resolution-limited sharp and well defined\cite{np13_117}. These results indicate disorder is critical in understanding the magnetic excitation spectra both at zero and high fields, similar to \ymgo~(Refs.~\onlinecite{PhysRevLett.118.107202,np13_117}). Given the absence of magnetic thermal conductivity and presence of frequency-dependent ac susceptibility\cite{PhysRevLett.120.087201}, and the critical role of disorder plays in the underlying physics, we believe \yzgo possesses a disorder-induced spin-glass ground state. Such a state can explain all the observations in this material so far, including the absence of static magnetic order, broad continuous magnetic excitation spectra, zero magnetic thermal conductivity, and especially the frequency-dependent freezing peaks in the ac susceptibility\cite{PhysRevLett.120.087201}.

%Because of the similarity between \yzgo and \ymgo, we believe it is reasonable to extend this conclusion to \ymgo as well\cite{PhysRevLett.120.087201,nc12_4949}.

Taking into account the disorder effect, there are some alternative ground states for \yzgo and \ymgo as well. For instance, the random-singlet state is a promising candidate\cite{PhysRevX.8.031028,PhysRevB.103.205122,
nc9_4367}. In such a state, two short-range antiparallel spins couple into a singlet. Without disorder, it is a valence-bond solid which breaks the rotational symmetry\cite{nature464_199}. By introducing the quenched bond disorder, the valence-bond-solid state is destroyed, and a random-singlet phase with nucleation of topological defects carrying spin-1/2 moments emerges. {\new These spins can produce an analogous spin-glass freezing phenomenon. Therefore, disordered spin-glass state may be regarded as the classical limit of a random-singlet state.} This proposal has many interesting consequences that are consistent with experimental observations. First, with the presence of disorder, the system does not approach magnetic phase transition down to ultralow temperatures. Second, when simulating the dynamical structure factor, the continuum-like spectra along edges of the Brillouin zones observed in INS experiments are also reproduced\cite{PhysRevX.8.031028}. Third, besides the INS results, this proposal also properly interprets the power-law behavior of the specific heat\cite{PhysRevX.8.031028}. Finally, the system can freeze into a spin glass as well in some cases, consistent with the ac susceptibility measurements\cite{PhysRevLett.120.087201,nc12_4949}.

For the sake of completeness, we wish to discuss several unusual features especially in \ymgo that may be at odds with aforementioned scenarios. First is the observation of persistent spin dynamics and absence of spin freezing in the $\mu$SR experiments on \ymgo~(Refs.~\onlinecite{PhysRevLett.117.097201,PhysRevB.102.014428}), which contradicts with the ac susceptibility on both \ymgo and \yzgo~(Refs.~\onlinecite{PhysRevLett.120.087201,nc12_4949}). One possible explanation will be that these two techniques probe different time windows of the spin dynamics, and the former and latter are sensitive to fast and slow spin fluctuations, respectively\cite{npjqm4_12}. Second, ultralow-temperature dc susceptibility measured with a Faraday force magnetometer does not reveal a bifurcation between the zero-field-cooling and filed-cooling curves at the freezing temperature as expected for a spin glass, although there is a clear kink around the freezing temperature\cite{PhysRevLett.122.137201,PhysRevLett.120.087201}. Third, and probably most interestingly, is a very recent thermal conductivity measurement on \ymgo~(Ref.~\onlinecite{nc12_4949}).  It shows zero residual term for $\kappa$ measured along the $c$-axis, consistent with previous conclusions on \ymgo and \yzgo~(Refs.~\onlinecite{PhysRevLett.117.267202,PhysRevLett.120.087201,
nc12_4949}). On the other hand, $\kappa_a/T$ ($\kappa_a$ is the thermal conductivity along the $a$-axis) shows a finite linear term of $\kappa_{a0}/T=0.0058$ or 0.0016~W~m$^{-1}$~K$^{−2}$ depending on the fitting range, indicative of the survival of magnetic excitations with mean-free path a few times of the spin-spin distance, even in the presence of disorder in this material\cite{nc12_4949}. This is in contrast to earlier reports where the in-plane $\kappa_{0}/T$ is essentially zero for both materials\cite{PhysRevLett.117.267202,PhysRevLett.120.087201}. While the discrepancy between these experiments is unclear at this time and remains to be resolved, disorder is considered to be responsible for suppressing the magnetic thermal conductivity in Ref.~\onlinecite{nc12_4949} as well. These results indicate that the spin-glass ground state in these materials is rather unusual and complex, and should deserve further investigations both from theory and experiment. The bottom line for these discussions, is perhaps that disorder is an important ingredient in the underlying physics and any proposed ground state should take it into account.

\section{Summary}

To summarize, we have performed INS measurements on high-quality \yzgo single crystals under a $c$-axis magnetic field to examine the field evolution of the magnetic excitations. A moderate filed of 2.5~T redistributes the spectral weight of the broad continuum at zero field in a more uniform fashion in the energy-momentum space. When applying a field up to 10~T which drives the system into a ferromagnetic state, we observe clear spin-wave excitations with a gap of $\sim$1.4 meV. However, different from the sharp and well-defined spin waves expected for the ferromagnetic state, the spectra exhibit strong broadening in energy and momentum. By considering the disorder effect, our classical Monte Carlo simulations can reproduce not only the continuous spectra at zero field, but also the broad ferromagnetic spin waves at 10-T field. These results demonstrate that disorder is an important parameter in shaping the spin dynamics as well as in governing the ground state in frustrated quantum magnetic systems.

\section{Acknowledgements}

The work was supported by National Key Projects for Research and Development of China with Grant No.~2021YFA1400400, the National Natural Science Foundation of China with Grant Nos.~11822405, 12074174, 12074175, 11774152, 11904170, 12004251, 12004249, and 12004191, Natural Science Foundation of Jiangsu Province with Grant Nos.~BK20180006, BK20190436, and BK20200738, Hubei Provincial Natural Science Foundation of China with Grant No.~2021CFB238, Shanghai Sailing Program with Grant Nos.~21YF1429200 and 20YF1430600, Fundamental Research Funds for the Central Universities with Grant No.~020414380183, and the Office of International Cooperation and Exchanges of Nanjing University. We would like to thank MLZ for allowing us to use the neutron Laue diffractometer NLaue for the coalignment of the \yzgo single crystals. We thank ILL for allowing our access to the neutron-scattering facility through Proposal No.~4-05-680~(Ref.~\onlinecite{ILL4-05-680}), and the support from the Sample Environment Group for setting up the dilution refrigerator on ThALES. Z.M. thanks Beijing National Laboratory for Condensed Matter Physics for funding support.

%\bibliography{topo}


\end{document}